\documentclass[%
 reprint,
superscriptaddress,
 amsmath,amssymb,
]{revtex4-2}

\usepackage{graphicx}
\usepackage{dcolumn}
\usepackage{bm}
\usepackage{braket}         
\usepackage{amsmath}
\usepackage{amsfonts}
\usepackage{bbold}          
\usepackage{subfigure}      
\usepackage{hyperref}       

\begin{document}

\title{Comparing the hardness of MAX 2-SAT problem instances for quantum and classical algorithms}

\author{Puya~Mirkarimi}
    \email{puya.mirkarimi@durham.ac.uk}
    \affiliation{Department of Physics, Durham University, Durham DH1 3LE, United Kingdom}
\author{Adam~Callison}
    \affiliation{Blackett Laboratory, Imperial College London, London SW7 2BW, United Kingdom}
    \affiliation{Department of Physics and Astronomy, University College London, London WC1E 6BT, United Kingdom}
\author{Lewis~Light}
    \affiliation{Department of Physics, Durham University, Durham DH1 3LE, United Kingdom}
\author{Nicholas~Chancellor}
    \affiliation{Department of Physics, Durham University, Durham DH1 3LE, United Kingdom}
\author{Viv~Kendon}
    \affiliation{Department of Physics, Durham University, Durham DH1 3LE, United Kingdom}
    \affiliation{Department of Physics, SUPA and University of Strathclyde, Glasgow G4 0NG, United Kingdom}
    
\date{July 21, 2023}

\begin{abstract}

An algorithm for a particular problem may find some instances of the problem easier and others harder to solve, even for a fixed input size.
We numerically analyse the relative hardness of MAX 2-SAT problem instances for various continuous-time quantum algorithms and a comparable classical algorithm.
This has two motivations: to investigate whether small-sized problem instances, which are commonly used in numerical simulations of quantum algorithms for benchmarking purposes, are a good representation of larger instances in terms of their hardness to solve, and to determine the applicability of continuous-time quantum algorithms in a portfolio approach, where we take advantage of the variation in the hardness of instances between different algorithms by running them in parallel.
We find that, while there are correlations in instance hardness between all of the algorithms considered, they appear weak enough that a portfolio approach would likely be desirable in practice.
Our results also show a widening range of hardness of randomly generated instances as the problem size is increased, which demonstrates both the difference in the distribution of hardness at small sizes and the value of a portfolio approach that can reduce the number of extremely hard instances.
We identify specific weaknesses of these quantum algorithms that can be overcome with a portfolio approach, such their inability to efficiently solve satisfiable instances (which is easy classically).
\end{abstract}

\maketitle

\section{\label{sec:introduction}Introduction}

Are the small-sized problem instances typically used for numerical simulations actually difficult enough to solve to provide a useful test of quantum algorithms?  
We investigate this question in the setting of continuous-time quantum computing (adiabatic quantum computing, quantum walks, and quantum annealing in particular) used to solve hard optimization problems.
Here, the word ``difficult'' refers to the amount of computing resources used to solve one particular instance of a problem, and ``hard'' refers to the scaling of the computational complexity with respect to input size.
Not all instances of hard problems are actually difficult to solve, even when they belong to a problem class that is NP-hard.  
Complexity classes are concerned with the asymptotic behaviour of computational complexity as a function of input size, and even for uniformly hard problem classes, there are instances that are much less difficult to solve than the others, although they may form a vanishingly small subset in the large size limit.
However, for the small sizes we have to use for numerical simulations, the less difficult instances could form a significant fraction of the instances being processed, and this could significantly skew the results of the simulations.

Despite the above caveats, prior work by some of the authors
\cite{Callison2019} found surprisingly good scaling for as few as five qubit spin glass ground state problems, using quantum walk computation.
This contrasts with the search problem \cite{Roland2002,Shenvi2003}, where finite size effects are apparent up to 20 or so qubit problem sizes \cite{Morley2017} for all continuous-time quantum computing methods.
Crosson et al.~\cite{Crosson2014} identified 20 qubit sized instances of MAX 2-SAT that are difficult for quantum annealing (low success probabilities at anneal time $t_f=100$).
However, this does not guarantee that these instances are also difficult for classical algorithms, or for other continuous-time quantum computing methods, such as quantum walk computation.

If different problem instances are more or less difficult for different algorithms, the best solution method can be a hybrid approach. Portfolio solvers for the Boolean satisfiability problem (SAT) are an example of this. These solvers take a set of different core solvers or different configurations of the same core solver, called a portfolio, and they run the solvers in parallel on different computing cores. Heuristics and machine learning techniques can be used to determine smart resource allocations that assign more computing cores to solvers that are likely to perform better, based on the features of an instance. Portfolio-based SAT solving was introduced in 2008 with ManySAT~\cite{Hamadi2009}, and solvers using the portfolio approach have outperformed all other solvers in the parallel track of recent SAT competitions~\cite{satcompetition}.

In this work, we consider the difficulty of the MAX 2-SAT instances from~\cite{Crosson2014}, which are difficult for coherent quantum annealing, for other quantum and classical algorithms. We compare these difficult instances with typical MAX 2-SAT instances, and also compare the performance of a good classical algorithm.
We then evaluate parallel approaches that combine two quantum algorithms and approaches that combine a quantum algorithm and a classical algorithm, to identify portfolio-based strategies that could outperform a single algorithm used alone.

The paper is organised as follows. In Sec.~\ref{sec:background}, we give an introduction to methods in continuous-time quantum computing and the MAX 2-SAT problem. In Sec.~\ref{sec:numerical_methods}, we outline the datasets and methods we have used in our numerical analysis. Then, we present the results of our work in Sec.~\ref{sec:results}, where we make comparisons between the difficulty of instances for different algorithms and study the behaviour of these algorithms on satisfiable instances, which are classically easy to solve. Finally, we give an overview of the results and present our conclusions in Sec.~\ref{sec:conclusions}.

\section{\label{sec:background}Background}

In this section, we introduce the definitions and concepts used in this work. This includes an overview of continuous-time quantum computing in the coherent regime, a description of the MAX 2-SAT problem, and a mapping of MAX 2-SAT to a problem Hamiltonian. Definitions given in this section will be used when quantifying the success of algorithms in later sections. The contents of this section are not new work. Rather, this section is intended to be a brief review of key concepts and prior work for readers that are new to this literature.

\subsection{\label{sec:continuous-time_quantum_computing}Continuous-time quantum computing}

Continuous-time quantum computing is a model for computing that offers an intuitive approach to solving combinatorial optimization problems on quantum hardware. In this approach, the problem is encoded in a problem Hamiltonian $H_{\mathrm{problem}}$ such that the ground state of $H_{\mathrm{problem}}$ corresponds to the desired solution. The computation is performed by initialising a set of qubits to a state $\ket{\psi(0)}$, applying a time-dependent Hamiltonian, and measuring the final state of the qubits after a time $t_f$ has passed. Assuming that the system stays in a fully coherent regime for the full length of the computation, the evolution of the system between the initialisation and measurement steps can be described by the Schr\"{o}dinger equation. Typically, the Hamiltonian is expressed in the form
\begin{equation}
    H(t) = A(t) H_{\mathrm{driver}} + B(t) H_{\mathrm{problem}},
    \label{eq:total_hamiltonian}
\end{equation}
where $A(t)$ and $B(t)$ are the control functions, which are generally time-dependent real numbers, and $H_{\mathrm{driver}}$ is a Hamiltonian that drives state transitions.

The continuous-time quantum walk (QW)~\cite{Farhi1998} is a form of fully coherent continuous-time quantum computing where the control functions are time-independent. The initial state is chosen to be the ground state of $H_{\mathrm{driver}}$, which is known in advance and is easy to prepare. The QW Hamiltonian is given by
\begin{equation}
    H_{\mathrm{QW}}(\gamma) = \gamma H_{\mathrm{driver}} + H_{\mathrm{problem}},
    \label{eq:qw_hamiltonian}
\end{equation}
where we set $B(t)=1$ and $\gamma = A(t)$ is called the hopping rate. QW is a quantum analogue of the classical continuous-time random walk, which is a stochastic process that describes the path of a walker as it takes random steps on a mathematical space. The hopping rate $\gamma$ can be interpreted as the probability per unit time that the walker will move to an adjacent site. For discussions of the connection between QW and other forms of continuous-time quantum computing, see \cite{Wong2016a,Morley2017,Callison2019}.

Another coherent form of continuous-time quantum computing is adiabatic quantum computing (AQC)~\cite{Farhi2000}. In AQC, the system is prepared in the ground state of $H_{\mathrm{driver}}$, and the Hamiltonian is slowly varied from $H_{\mathrm{driver}}$ to $H_{\mathrm{problem}}$ by varying the control functions from $A(0) = 1$ and $B(0) = 0$ to $A(t_f) = 0$ and $B(t_f) = 1$. If $A(t)$ and $B(t)$ are smoothly and monotonically decreasing and increasing respectively and the minimum gap between the energy of the ground state and the first excited state is non-zero, then the adiabatic theorem~\cite{Born1928} ensures that the system will have a high probability of staying in the instantaneous ground state throughout the computation, provided that $t_f$ is long enough. In AQC, $t_f$ is always long enough to be close to the adiabatic limit. We refer to protocols with time-dependent control functions where coherence and/or adiabaticity are not guaranteed as quantum annealing (QA). Note that AQC and QA are sometimes defined differently elsewhere in the literature.

In practice, the problem Hamiltonian is typically expressed as an Ising Hamiltonian on $n$ qubits, which takes the form
\begin{equation}
    H_{\mathrm{problem}} = \sum_{i=1}^{n-1} \sum_{j=i+1}^{n} J_{ij} \sigma_i^z \sigma_j^z + \sum_{i=1}^{n} h_i \sigma_i^z,
    \label{eq:problem_hamiltonian_general}
\end{equation}
where the couplings $J_{ij} \in \mathbb{R}$ and field strengths $h_{i} \in \mathbb{R}$ are used to encode the problem, and $\sigma_i^z = \mathbb{1}_2^{\otimes i - 1} \otimes \sigma_z \otimes \mathbb{1}_2^{\otimes n - i}$ is the Pauli operator $\sigma_z$ acting on qubit $i$ and identities acting on all other qubits. A common choice for the driver Hamiltonian in both QW and AQC is the transverse-field Hamiltonian
\begin{equation}
    H_{\mathrm{driver}} = - \sum_{i=1}^{n} \sigma_i^x,
    \label{eq:driver_hamiltonian}
\end{equation}
where $\sigma_i^x$ is defined similarly to $\sigma_i^z$ as $\sigma_i^x = \mathbb{1}_2^{\otimes i - 1} \otimes \sigma_x \otimes \mathbb{1}_2^{\otimes n - i}$. The ground state of this driver Hamiltonian, which in QW and AQC is the initial state of the system, is the equal superposition of the computational basis states,
\begin{equation}
    \ket{\psi(0)} = \frac{1}{\sqrt{2^n}}\sum_{j=0}^{2^n-1} \ket{j} = \ket{+}^{\otimes n},
    \label{eq:ground_state_driver_hamiltonian}
\end{equation}
where $\ket{+}=(\ket{0}+\ket{1})/\sqrt{2}$.

Whereas in AQC the adiabatic theorem guarantees that one can always attain a near-unity probability of successfully measuring the state corresponding to the optimal solution by using a long enough run time, this is not the case for QW. Therefore, to find the optimal solution with a probability of close to unity using QW, one should perform the computation many times and take the best found solution. For simplicity, in the rest of this work we will only be considering problem instances that have a unique optimal solution corresponding to a non-degenerate ground state $\Ket{\psi_G}$ of the problem Hamiltonian. According to the Born rule~\cite{Griffiths2004Introduction}, the probability of measuring the state $\Ket{\psi_G}$ after evolving the system for a time $t_f$ is
\begin{equation}
    P(t_f) = \left| \Braket{\psi_G|\psi(t_f)} \right| ^2.
    \label{eq:success_probability}
\end{equation}
In general, the success probability $P(t)$ for QW fluctuates with time in an unpredictable manner. To avoid taking every measurement at a time when $P(t_f)$ happens to be near a local minimum, it is beneficial to use different values of $t_f$ for each measurement. We will consider the same approach as in~\cite{Callison2019}, where $t_f$ is selected uniformly at random from an interval $I=[t_I, t_I + \Delta t_I]$. The average single run success probability is defined as
\begin{equation}
    \overline{P}(t_I, \Delta t_I) \equiv \frac{1}{\Delta t_I} \int_{t_I}^{t_I + \Delta t_I}  P(t_f) \, dt_f,
    \label{eq:average_success_probability_qw}
\end{equation}
which is the mean success probability of individual measurements in this approach. The number of repeats required to attain an arbitrarily high probability of measuring the ground state scales as the inverse of $\overline{P}(t_I, \Delta t_I)$ for small $\overline{P}(t_I, \Delta t_I)$.

For both QW and AQC, the choice of control functions is a free parameter that can affect the performance of the algorithms. In QW, this corresponds to the hopping rate $\gamma$. A good choice for $\gamma$ is one that balances the energy between $H_\mathrm{driver}$ and $H_\mathrm{problem}$ in the total Hamiltonian~\cite{Callison2019}. To achieve this, we would like to set $\gamma$ such that the energy-spread of $\gamma H_\mathrm{driver}$ is equal to the energy-spread of $H_\mathrm{problem}$. For the transverse-field driver Hamiltonian defined in Eq.~\eqref{eq:driver_hamiltonian}, the energy-spread is $2n$. However, the energy-spread of the problem Hamiltonian, which is the difference between the maximum and minimum number of clauses that can be satisfied, depends on the particular problem instance and it is not possible to calculate it without solving the instance. Therefore, we instead consider the average energy-spread of $H_\textrm{problem}$ for instances of a given number of variables $n$, and we calculate a heuristic hopping rate $\gamma_\textrm{heur}$ by setting this equal to the energy-spread of $\gamma H_\textrm{driver}$, giving
\begin{equation}
    \gamma_\textrm{heur} = \frac{\langle E_{2^n} - E_1 \rangle}{2n}.
    \label{eq:heuristic_hopping_rate}
\end{equation}
Here, $E_1$ and $E_{2^n}$ are the smallest and largest eigenvalues of $H_\textrm{problem}$ respectively.
For our analysis, the average energy-spread $\langle E_{2^n} - E_1 \rangle$ was calculated for each $n$ by diagonalising the problem Hamiltonians of the generated instances (i.e. solving the problems). In practice, when the size of the instances makes this approach too computationally expensive, $\langle E_{2^n} - E_1 \rangle$ can be calculated for similar instances with fewer variables and extrapolated to larger $n$.

The choice of control functions that we have used for all AQC simulations in this analysis is the linear schedule
\begin{equation}
    A(t) = 1 - \frac{t}{t_f}, ~ B(t) = \frac{t}{t_f}.
    \label{eq:aqc_control_functions}
\end{equation}
While there exist strategies involving QW and AQC with different choices of control functions that can improve performance~\cite{callison2021energetic}, we will not be exploring them in this work. The simulations in~\cite{Crosson2014} used the same linear schedule as above with a constant duration $t_f=100$, which is close to the adiabatic limit for most of the instances that were generated in their work. (Over half of the instances had success probabilities of $P(100) > 0.95$.) However, the instances that were selected for being difficult had success probabilities of $P(100) < 10^{-4}$, which puts their simulations far from the adiabatic limit. Hence, these instances are difficult for a coherent QA protocol that relaxes the condition of adiabaticity in AQC.

\subsection{\label{sec:max_2-sat}MAX 2-SAT}

A Boolean formula $\phi = \phi(x_1, \dots , x_n)$ consists of $n$ Boolean variables $x_1, \dots , x_n$, Boolean operators, and parentheses. The Boolean operators we will consider are conjunction ($\land$), disjunction ($\lor$), and negation ($\lnot$). Boolean variables can take one of the two possible logical values true (denoted $0$) and false (denoted $1$). A set of values that are assigned to the $n$ variables in a formula is called an assignment, and for each assignment the Boolean formula $\phi$ will evaluate to either true or false.
We define $2n$ literals $l_1, \dots, l_{n}$ and $l_{-n}, \dots, l_{-1}$ such that the literal $l_i$ is associated with the variable $x_i$ if $i$ is positive, or $\neg x_{|i|}$ if $i$ is negative.

In the maximum satisfiability problem (MAX SAT), a problem instance is specified by a Boolean formula $\phi$ that is in conjunctive normal form (CNF), which is a formula that is structured as a conjunction of $m$ clauses, where a clause is a disjunction of literals. In this work, we will be studying maximum 2-satisfiability (MAX 2-SAT), which is a special case of MAX SAT where there are two literals in each clause. An example of a valid formula for MAX 2-SAT is
\begin{equation}
\begin{gathered}
    (x_1 \lor x_2)\land(\neg x_1 \lor x_2)\land(x_1 \lor \neg x_3)\,\land \\
    (\neg x_1 \lor x_3)\land(\neg x_2 \lor x_3)\land(\neg x_2 \lor \neg x_3),
    \label{eq:cnf_example}
\end{gathered}
\end{equation}
where $n=3$ and $m=6$ in this case. In MAX 2-SAT, any possible truth assignment is known as a solution, and we are tasked with finding an optimal solution, which is a solution that maximises the number of clauses that evaluate to true. A clause that evaluates to true is said to be \textit{satisfied}, and a clause that evaluates to false is said to be \textit{unsatisfied}. The formula above has four optimal solutions which each satisfy five of the six clauses, one of which is the assignment $x_1 = 0,~x_2 = 0,~x_3 = 0$.

Although 2-SAT, the decision version of MAX 2-SAT, is in the complexity class P~\cite{Aspvall1979}, MAX 2-SAT is NP-hard~\cite{Garey1974}. Nevertheless, MAX SAT solvers have made remarkable advancements over the past three decades, and they are able to solve or approximately solve MAX 2-SAT instances with relatively large input sizes. This progress can in part be attributed to annual competitions in producing the fastest SAT and MAX SAT solvers~\cite{maxsatevaluations, satcompetition} and the large demand for these solvers, which is generated from the ability to efficiently map a wide range of practical problems to satisfiability problems. Examples include integrated circuit design debugging~\cite{safarpour2007improved, chen2010automated}, cancer therapy design~\cite{Lin2012}, software verification~\cite{Clarke2004, Ivancic2008}, and planning~\cite{Kautz1992, Rintanen2012}.

Outside the title and abstract of this paper, we avoid using the word ``hardness'' unless we are referring to the scaling of the computational complexity of a problem. When referring to the amount of resources used by a given algorithm to solve a particular instance of a problem, we use the word ``difficulty'' instead. Note that some papers use the word ``hardness'' instead of ``difficulty'', which is what we have done in the title and abstract. An instance that is difficult for one algorithm may not necessarily be difficult for another algorithm. The hardness of a problem is typically measured by the worst-case or average-case computational complexity. Due to MAX 2-SAT being NP-hard~\cite{Garey1974}, we expect a worse than polynomial scaling of the worst-case time complexity with $n$ for any algorithm, assuming $\textrm{P} \neq \textrm{NP}$.

In practice, the average run time scaling for a given set of instances may differ significantly from the worst-case time complexity of the problem and may be polynomial, even if the problem is NP-hard. The way in which instances are sampled is often an important consideration when performing an analysis on run times, especially when the problem is not uniformly hard. It has been shown that the difficulty of random MAX 2-SAT instances increases with the clause density $\rho=m/n$, and that there is a difficulty phase transition at the critical clause density $\rho_c = 1$~\cite{Coppersmith2004, FernandezdelaVega2001}. This has been demonstrated experimentally for QA~\cite{Santra2014} and has also been observed in numerical simulations of the quantum approximate optimisation algorithm~\cite{Akshay2020, Golden2023}.

\subsection{\label{sec:problem_mapping}Problem mapping}

In order to solve instances of MAX 2-SAT with a continuous-time quantum algorithm, a mapping of the problem as a Hamiltonian in the form given by Eq.~\eqref{eq:problem_hamiltonian_general} is required. Such a mapping should assign lower energies to eigenstates corresponding to more desirable solutions. Under the binary encoding of Boolean variables $x_i \in \{0, 1\}$ where 0 corresponds to true and 1 corresponds to false, the disjunction operator is equivalent to multiplication---i.e.\ $x_i \lor x_j$ can be written as $x_i x_j$. By identifying the variable $x_i$ with the single-qubit basis state $\Ket{x_i}$, we observe that
\begin{equation}
    \frac{ \mathbb{1} - \sigma_z }{2} \Ket{x_i} = x_i \Ket{x_i},
    \label{eq:binary_operator}
\end{equation}
where $\sigma_z$ is the Pauli Z operator. For a clause $C_k = l_i \lor l_j$, where the literal $l_i$ is positive (negative) if the number $i$ is positive (negative), the corresponding term in the problem Hamiltonian can be constructed by taking the product
\begin{equation}
    H_{C_k} = \frac{\mathbb{1} - \mathrm{sgn}(i) \sigma_{|i|}^z}{2} \frac{\mathbb{1} - \mathrm{sgn}(j) \sigma_{|j|}^z}{2},
    \label{eq:clause_hamiltonian}
\end{equation}
where we have used the sign function $\mathrm{sgn}$ to extract the sign of the indices. This term contributes an energy equal to $x_i x_j$. The problem Hamiltonian can then be constructed by taking a sum over the terms corresponding to each of the clauses in the Boolean formula $\phi$, giving
\begin{equation}
    H_\mathrm{problem} = \sum_{C_k \in \phi} H_{C_k}.
    \label{eq:max2sat_hamiltonian}
\end{equation}
The eigenvalues of this Hamiltonian are equal to the numbers of clauses that are unsatisfied by the assignments corresponding to the eigenstates.

\subsection{\label{sec:algorithm_portfolios}Algorithm portfolios}

One of the aims of this paper is to investigate the extent to which a portfolio-based strategy could improve the performance of continuous-time quantum algorithms. The portfolio approach is a simple method of achieving parallelism that was inspired by strategies for managing risk while increasing utility in economics~\cite{Huberman1997}. It takes advantage of the lack of correlation in the difficulty of instances between several algorithms (together called a portfolio) by running the different algorithms in parallel. This approach has been applied to SAT solving~\cite{Hamadi2009, Balyo2015}, where it typically outperforms all other parallel strategies in competitions~\cite{satcompetition}. Optimal portfolios of recent SAT solvers and the impact of the portfolio size on performance have been studied~\cite{Bach2022}. For overviews of classical parallel SAT solving, see~\cite{Martins2012, Holldobler2011}.

An advantage of the portfolio approach is that it has the potential to decrease a strategy's sensitivity to extremely difficult instances. As a simple example of this, consider two algorithms that each find a subset of instances extremely difficult to solve, where these two subsets do not overlap. These instances may not only significantly impact the worst-case performance of the algorithms but also the mean performance. If these algorithms were combined into a portfolio---for example, by running the two algorithms in parallel and allocating each algorithm half of the computing resources---the instances that are extremely difficult for each algorithm would be solved more efficiently by the other algorithm. This speedup comes at the cost of decreased performance for instances that the two algorithms find similarly difficult. Similar discussions of such performance/sensitivity trade-offs have been made in the context of no free lunch theorems for optimisation~\cite{Wolpert1995, Wolpert1997, Ho2002}. While the no free lunch theorems do not apply to MAX 2-SAT in particular~\cite{McDermott2020}, an analysis of the implications of these theorems for portfolios of quantum and classical algorithms would be an interesting direction for future work.

In the context of quantum computing, parallel computing and the portfolio approach in particular have not been studied in significant depth. In~\cite{Maurer2001}, classical and quantum portfolios of quantum algorithms were found to perform better than standalone quantum algorithms for random 3-SAT. In this paper, we build on these results by studying the practical implications of a classical portfolio-based strategy for a selection of quantum and classical algorithms for MAX 2-SAT.

In related works that aren't specifically on the subject of the portfolio approach, various hybrid quantum-classical algorithms for near-term quantum devices have been studied~\cite{Callison2022}. There have been efforts towards integrating quantum processors into modern high performance computing systems~\cite{Humble2021}, which is an important step towards practically implementing a portfolio of quantum and classical algorithms. With regards to parallelism in continuous-time quantum computing, Pelofske et al. studied the use of quantum annealers to solve multiple independent problems in parallel on a single device~\cite{Pelofske2022, Pelofske2022Quantum}.

\section{\label{sec:numerical_methods}Numerical methods}

\subsection{\label{sec:datasets}Datasets}

We performed our numerical study on two sets of MAX 2-SAT instances. The first set of instances were generated by Crosson et al. in~\cite{Crosson2014}, and each instance in this set contains $n=20$ Boolean variables and $m=60$ unique clauses.
Having a constant clause density $\rho=3$ that is well above the critical clause density ensures that these instances are in the difficult regime. Each of the clauses in these instances were generated by randomly selecting two literals that are associated with distinct variables. Instances with multiple optimal solutions (corresponding to degenerate ground states of the problem Hamiltonian) were discarded.
Note that this may lead to a different relation between instance difficulty and clause density compared to the phase transition results discussed in Sec.~\ref{sec:max_2-sat} for random MAX 2-SAT. We are not aware of any results related to the phase transition in the context of instances with unique optimal solutions.
202,078 of such instances were generated, but only those with an AQC success probability at time $t_f=100$ of $P(100) < 10^{-4}$ were selected, meaning that the 137 instances that remained are difficult for QA. For convenience, these instances were transformed by negating all literals corresponding to variables that were set to $1$ in the original optimal assignment so that the optimal solution is always the $00\dots0$ bit string.

The second set of instances were generated in a similar manner as those from~\cite{Crosson2014}. For each number of variables in the range $5 \leq n \leq 20$, 10,000 instances containing $m=3n$ unique clauses were generated with randomly selected pairs of literals corresponding to distinct variables for each clause, and only the instances with unique optimal solutions were kept. The transformation to set $00\dots0$ as the optimal solution was applied. Unlike the instances from~\cite{Crosson2014}, there was no post-selection of a fraction of these instances based on difficulty; hence, we will refer to these as the ``typical'' instances. It has been shown that instances with an unbalanced ratio of positive to negative literals (or unnegated to negated variables) may be more difficult than balanced instances, where there are an equal number of positive and negative literals~\cite{Austrin2007}. These instances are balanced on average and hence aren't maximally difficult.

\subsection{\label{sec:numerical_tests}Numerical tests}

All of the findings in this paper are results of numerical simulations that were carried out using the Python programming language~\cite{VanRossum1995}. The NumPy~\cite{harris2020array} and Scipy~\cite{2020SciPy-NMeth} libraries were used for computationally intensive calculations, and matplotlib~\cite{hunter2007matplotlib} was used for plotting. The implementation of the least-squares method in scipy.optimize.curve\_fit was used for obtaining the linear fits in this paper. The unitary time-evolutions of quantum systems were simulated using quimb~\cite{Gray2018} as a convenient interface to scipy.integrate.complex\_ode and its implementation of dop853~\cite{Hairer1993}, which is a Runge-Kutta method of order 8 with an adaptive step-size. To calculate the average success probabilities in the limit of infinite time interval (shown in the Appendix), Hamiltonians were diagonalised using numpy.linalg.eigh. Simulations of QW were run on high performance computers at Imperial College London, and simulations of AQC were run on the Hamilton high performance computing cluster at Durham University.

\begin{figure}
    \centering
    \includegraphics[width=\columnwidth]{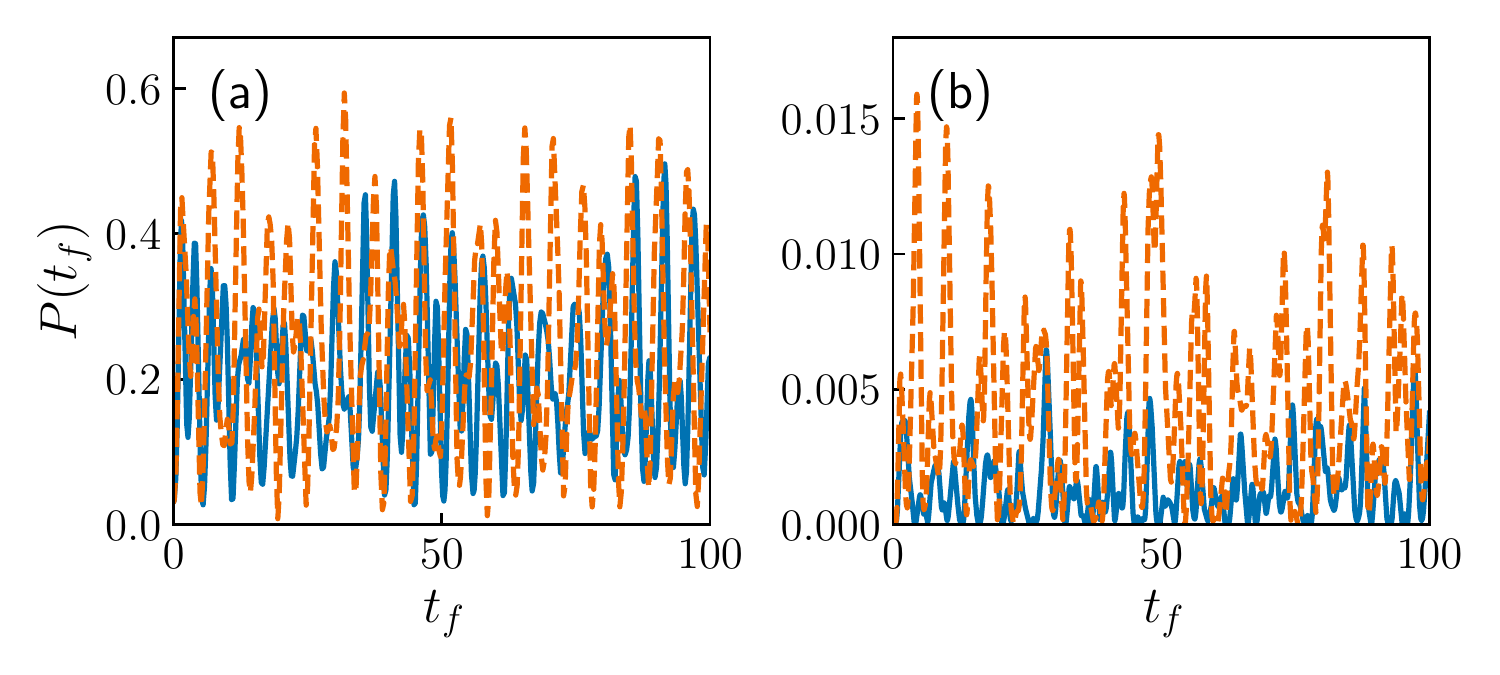}
    \caption{
    Instantaneous QW success probability $P(t_f)$ for two randomly selected pairs of instances (solid blue and dashed orange lines) with (a) $n=5$ variables and (b) $n=20$ variables plotted against the measurement time $t_f$. Note the difference in the upper limits of the y-axes.
    }
    \label{fig:m2sinstsuccessprobs}
\end{figure}

The quantity of interest for our analysis of QW is the average single run success probability $\overline{P}(t_I, \Delta t_I)$ given in Eq.~\eqref{eq:average_success_probability_qw}, which is defined over an interval of measurement times $I = [t_I, t_I + \Delta t_I]$. We set $t_I = 0$ and $\Delta t_I = 100$ for our calculations, which produces an interval that is longer than the timescale of the QW dynamics.
To demonstrate this, we plot the instantaneous QW success probability in this interval for pairs of randomly selected instances with $n=5$ and $n=20$ variables in Fig.~\ref{fig:m2sinstsuccessprobs}, and we see that there are many oscillations in the success probability within the time interval in each case.
As shown in the Appendix, the specific choice of $\Delta t_I$ does not significantly impact the results as long as it is longer than the timescale of the QW dynamics.
The average success probability $\overline{P}(0, 100)$ was approximated by numerically integrating the Schr\"{o}dinger equation over the full time interval and taking a weighted average of the instantaneous success probabilities of the solutions that were evaluated at each iteration of the integration method, where the weights are given by the amount of time between successive iterations.

For AQC, the quantity of interest is the minimum evolution duration required to achieve a 99\% success probability $t_{0.99}$.
To find $t_{0.99}$ for a given instance, the quantum dynamics were simulated for a small duration $t_f$ and the success probability was calculated according to Eq.~\eqref{eq:success_probability}. If the success probability was less than 99\%, $t_f$ was doubled and a new success probability was calculated. This was repeated until either a success probability of greater than 99\% was found or the simulations became too computationally intensive to continue, in which case $t_{0.99}$ was not found. If the original success probability was greater than 99\%, $t_f$ was instead halved each time until a duration with less than 99\% success probability was found. The bisection method was then used to search for $t_{0.99}$ within the interval of the last two durations until a precision of at least 1\% was reached.

The classical algorithm MixBandB~\cite{Callison2021} was applied to all of the generated MAX 2-SAT instances to compare its performance against the quantum algorithms.
MixBandB was written to mirror some of the key characteristics of a highly competitive MAX SAT solver known as MIXSAT~\cite{Wang2019}, without including many of the heuristic methods that MIXSAT employs.
Like MIXSAT, MixBandB is a branch-and-bound algorithm that uses the ``Mixing method''~\cite{Wang2017} as a semidefinite programming solver in order to produce lower bounds, and rounding to produce good guesses.
MixBandB does not use a dual initialisation strategy or any of the data structure or implementation optimizations that are present in MIXSAT.
For each instance, we measured the number of times MixBandB accessed the problem specification, which we refer to as the number of problem calls $N_\mathrm{calls}$. This quantity serves as a proxy for the run time of the algorithm.

\section{\label{sec:results}Results}

In this section, we present an analysis of the difficulty of MAX 2-SAT instances for various quantum and classical algorithms based on the results of numerical simulation.
The QW success probability $\overline{P}(0, 100)$ is used as a measure of the difficulty of instances for QW, where a higher success probability corresponds to a less difficult instance, and the AQC duration $t_{0.99}$ is used as a measure of difficulty of instances for AQC, where a longer duration corresponds to a more difficult instance. We start by making a cross-comparison between the difficulty of MAX 2-SAT instances for QW and AQC
in Sec.~\ref{sec:QW/AQC_difficulty}, and we make further comparisons that include difficulty for MixBandB in Sec.~\ref{sec:quantum/classical_difficulty}.
Where we were able to obtain results for 20-variable instances, a comparison with QA difficulty using the instances from~\cite{Crosson2014} is also made.
In Sec.~\ref{sec:satisfiable_instances}, we investigate the relative difficulty of satisfiable instances, which are classically easy to solve, for the algorithms that we are considering.

\subsection{\label{sec:QW/AQC_difficulty}QW/AQC difficulty comparison}

\begin{figure}
    \centering
    \includegraphics[width=\columnwidth]{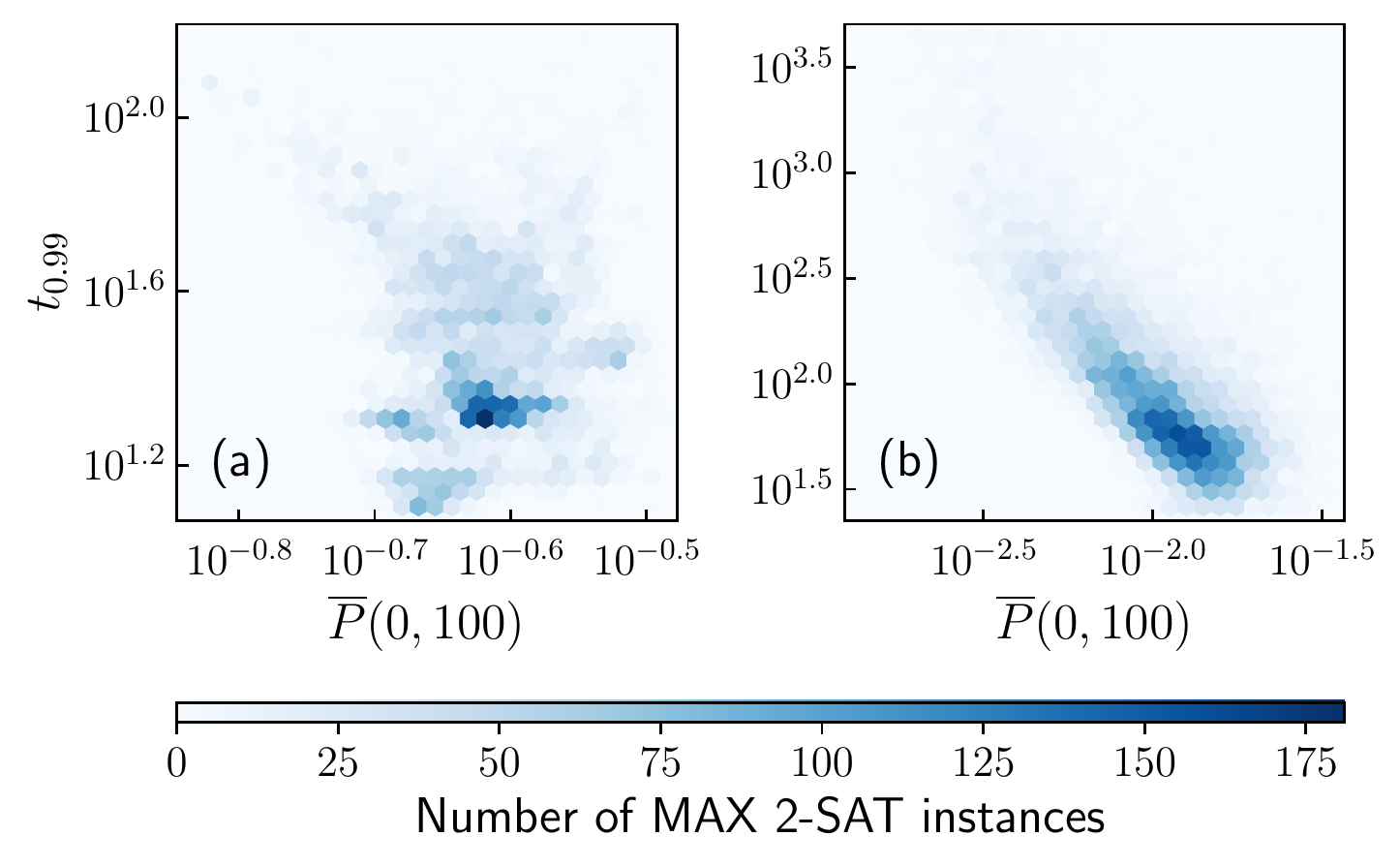}
    \caption{
    Heatmaps of the AQC duration $t_{0.99}$ against the QW success probability $\overline{P}(0,100)$ for typical MAX 2-SAT instances with (a) $n=5$ variables and (b) $n=15$ variables, excluding 138 instances of size $n=15$ for which $t_{0.99}$ was not successfully calculated. Visually, the plot in (b) looks reasonably well correlated, whereas the plot in (a) does not appear to show much correlation. This is confirmed by the Spearman's rank correlation coefficients, which were calculated to be $\approx 0.01$ for $n=5$ and $\approx -0.76$ for $n=15$.
    }
    \label{fig:m2saqcqwhexbin}
\end{figure}

To characterise the relation between the QW difficulty and AQC difficulty of MAX 2-SAT instances, Fig.~\ref{fig:m2saqcqwhexbin} shows the joint distribution of $\overline{P}(0, 100)$ and $t_{0.99}$ for typical instances with $n=5$ variables and for $n=15$ variables---the latter being the largest problem size that we could calculate $t_{0.99}$ for. We calculate the Spearman's rank correlation coefficient for the distributions to be $\approx 0.01$ and $\approx -0.76$ for $n=5$ and $n=15$ respectively. The Spearman's rank correlation coefficient has a range of -1 (perfect anti-correlation between the rankings of the two quantities) through 0 (no correlation in rankings) to +1 (perfect correlation in rankings).  Recalling that smaller $P(0, 100)$ and larger $t_{0.99}$ values indicate more difficult instances, the increasing negative Spearman's rank correlation coefficient values (anti-correlation) indicates a correlation between QW and AQC difficulty that gets stronger with $n$. This suggests that while a portfolio-based strategy may be able reduce the total run time, it would not produce a huge improvement at higher $n$, as the instances that are difficult for one algorithm aren't likely to be found less difficult by the other algorithm.

In Fig.~\ref{fig:m2saqcqwhexbin}(b), a long tail of instances that are extremely difficult for both QW and AQC can be identified visually. These instances are towards the top-left of the heatmap and are far from the location where the heatmap shows the highest density of instances. In comparison, the tail of instances to the bottom-right of the heatmap is much shorter, indicating that the least difficult instances aren't as outlying as the most difficult instances for both QW and AQC. It is possible that typical instances of the $n=15$ problem size are more dominated by less difficult instances than would be found at larger problem sizes. In other words, the top and left sides of the graph may have a higher density of instances when plotted for larger $n$, due to there being more instances in the ``difficult tail'' of the distribution. If this is the case, then the most difficult of the typical instances may be a better representation of the types of instances that are typically found at larger problem sizes. On top of this, instances of practical interest may have a significantly different composition of more and less difficult instances than our sample of randomly generated instances. Therefore, it would be useful to find out whether the level of correlation between $\overline{P}(0, 100)$ and $t_{0.99}$ changes with the difficulty of the instances, but this is not easy to tell from these heatmaps.

A more detailed analysis of the difficulty of instances is required, to distinguish between the increase in difficulty with $n$ that is simply due to the increase in problem size, from the change in the distribution of the difficulty of the instances for QW and AQC at each $n$.
This is not easy to achieve because we expect the proportion of very difficult instances to increase with $n$.
To accommodate such an analysis, we have followed a similar approach to other authors who have partitioned instances according to a measure of their difficulty~\cite{Wecker2016, Ebadi2022a}.
Specifically, we have grouped the typical instances of each number of variables $n$ into deciles that are ranked by difficulty. Decile 1 contains the 10\% of the instances that are least difficult, decile 2 contains the next least difficult 10\% of the instances, and so on, with decile 10 containing the most difficult 10\% of the instances. This partitioning is done using the average QW success probability $\overline{P}(0, 100)$ as the measure of difficulty to produce ``QW difficulty deciles'' and similarly done using the 99\% success probability duration $t_{0.99}$ for AQC to produce ``AQC difficulty deciles''. For AQC, it is assumed that the instances that we were not able calculate $t_{0.99}$ for in a reasonable amount of time are the most difficult instances. To identify the extremely difficult instances, we have also grouped together the most difficult 1\% of instances at each $n$ for QW and for AQC using the same measures of difficulty as for the deciles. By defining the QW/AQC ``difficulty percentiles'' in a similar way as the difficulty deciles, we can refer to the instances that are on the boundaries of the deciles by their percentiles. For example, the most difficult instance in the 3rd QW difficulty decile is the 30th percentile instance for QW difficulty. Similarly, the most difficult instance for AQC that is not one of the most difficult 1\% of instances for AQC is the 99th percentile instance for AQC difficulty.

\begin{figure}
\centering
    \includegraphics[width=\columnwidth]{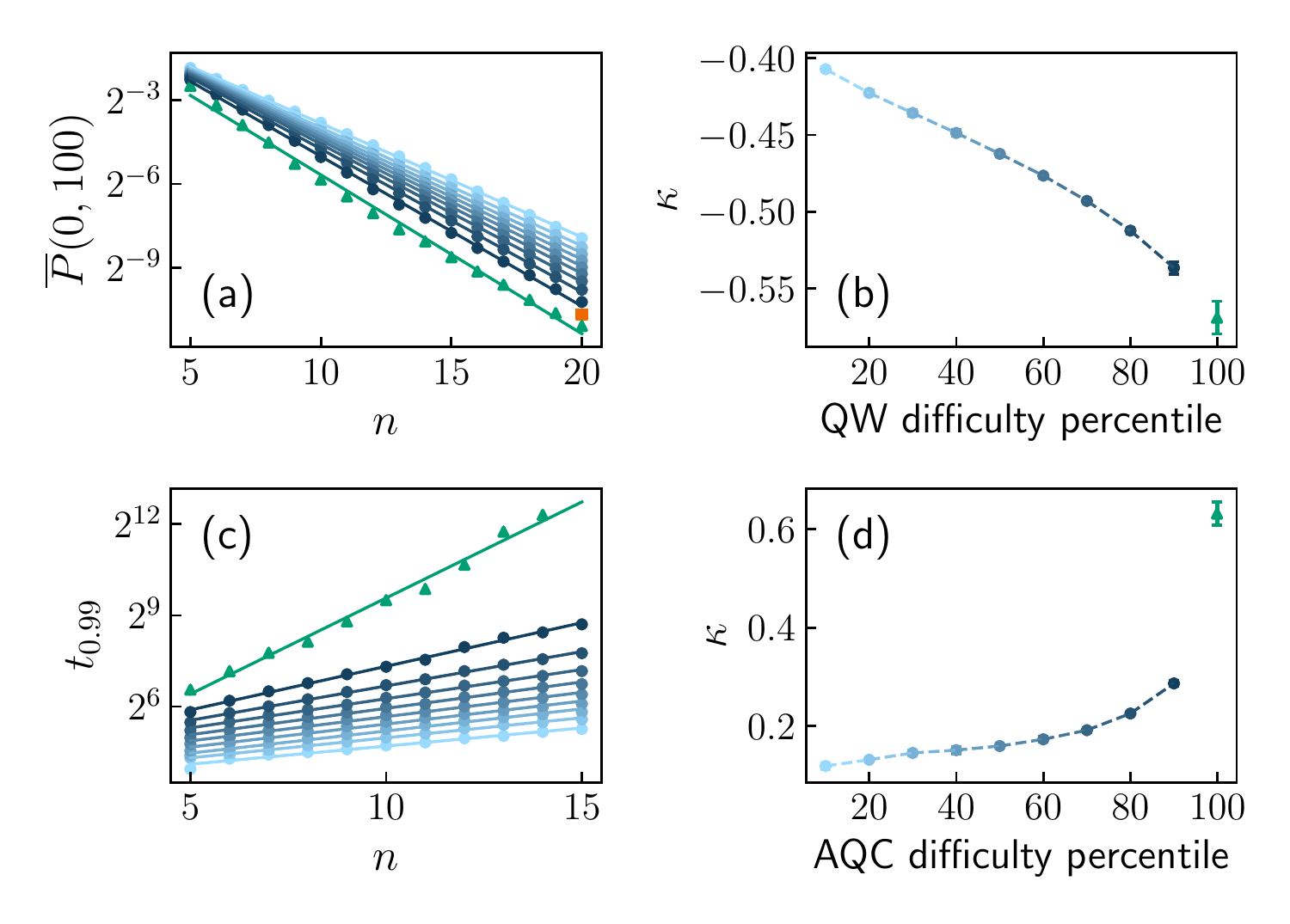}
    \caption{
    (a) The QW success probabilities $\overline{P}(0, 100)$ of the 10th, 20th, $\dots$, 90th percentile instances (blue circles) and the 99th percentile instances (green triangles) for QW difficulty plotted against the number of variables $n$ on log-linear axes, with a linear fit (solid line) for each percentile. The median QW success probability of the instances from~\cite{Crosson2014} is also plotted (orange square). Darker shades of blue represent more difficult percentiles. (b) Plot of the values of the scaling exponents $\kappa$ that have been inferred from the gradients of the linear fits, with error bars indicating standard errors. (c) and (d) Similar plots for the AQC duration $t_{0.99}$ and AQC difficulty percentiles. The results go up to $n=20$ for QW and $n=15$ for AQC due to time and resource constraints. Hence, there is no data point for the instances from~\cite{Crosson2014} shown in (c). Also note that we could not calculate $t_{0.99}$ for the 99th percentile instance for AQC difficulty at $n=15$, so there is no corresponding point in (c).
    }
    \label{fig:qw and aqc decile boundaries}
\end{figure}

Fig.~\ref{fig:qw and aqc decile boundaries}(a) shows a log-linear plot of the average success probabilities $\overline{P}(0, 100)$ for the most difficult instances of each QW difficulty decile (excluding the most difficult decile) and the 99th percentile instances for QW difficulty against the number of variables $n$. A linear fit is shown for each of the decile boundary instances and the 99th percentile instances, and the corresponding scaling exponents for each fit are plotted in Fig.~\ref{fig:qw and aqc decile boundaries}(b). These plots show that the scaling of $\overline{P}(0, 100)$ gets progressively worse as the subset of selected instances gets more difficult. Therefore, at larger problem sizes we can expect a bigger difference between the difficulty of the most and least difficult instances. The inferred scaling exponents also indicate that the tail of difficult instances gets longer as $n$ is increased, which means that the time spent solving many instances would be largely dominated by the most difficult instances. This highlights the value of a portfolio approach, as any efficiency improvement for the most difficult instances would make a significant difference to the total run time.

An orange point indicating the median value of $\overline{P}(0, 100)$ for the instances from~\cite{Crosson2014} is shown in Fig.~\ref{fig:qw and aqc decile boundaries}(a). The placement of this point shows that these instances are also difficult for QW, which suggests that there is a correlation between QW and QA difficulty. These instances were selected to be the most difficult 137 instances for QA out of 202,078 randomly generated instances, meaning that they are all in the top 0.1\% of the most difficult instances for QA. However, the median QW difficulty of these instance lies between the 90th and 99th percentiles of the typical instances, which indicates that the instances from~\cite{Crosson2014} are not as extremely difficult for QW as they are for QA. Given that these instances are at least four orders of magnitude more difficult for QA than the median of the randomly generated instances, their lower relative difficulty for QW is substantial, and an approach involving QA would benefit from speeding up these instances by running QW in parallel.

Figs.~\ref{fig:qw and aqc decile boundaries}(c) and~\ref{fig:qw and aqc decile boundaries}(d) show similar plots as above, but this time for the AQC duration $t_{0.99}$ and AQC difficulty deciles/percentiles. The median value of $t_{0.99}$ for the instances from~\cite{Crosson2014} was not calculated, as these instances were too large to run AQC simulations for. Just as with the QW results, we find that more difficult percentiles scale more harshly for AQC, and this effect is even more prominent than for QW. (The decile boundary scaling exponents range from $\kappa = 0.119 \pm 0.006$ to $\kappa = 0.286 \pm 0.005$ for AQC, as opposed to $\kappa = -0.407 \pm 0.002$ to $\kappa = -0.537 \pm 0.004$ for QW.) Notably, there is a large jump in $\kappa$ to $\approx 0.63\pm0.02$ between the 90th and 99th percentile instances, compared with a smaller jump to $\approx -0.57\pm0.01$ for QW, which implies that a small subset of the instances we are considering are extremely difficult for AQC. This is not surprising, as it is known that AQC performs very poorly on instances of other NP-hard problems when the minimum energy gap between the ground and first excited states is extremely small~\cite{Steiger2015}. The steep gradient of the fit for the 99th percentile instances may be an indication that the number of extremely difficult instances is growing with $n$, which would be consistent with the idea that we would see a larger fraction of instances in the ``difficult tail'' at larger $n$. Therefore, we are further motivated to analyse the relation between QW and AQC difficulty for these extremely difficult instances.

\begin{figure}
\centering
    \includegraphics[width=\columnwidth]{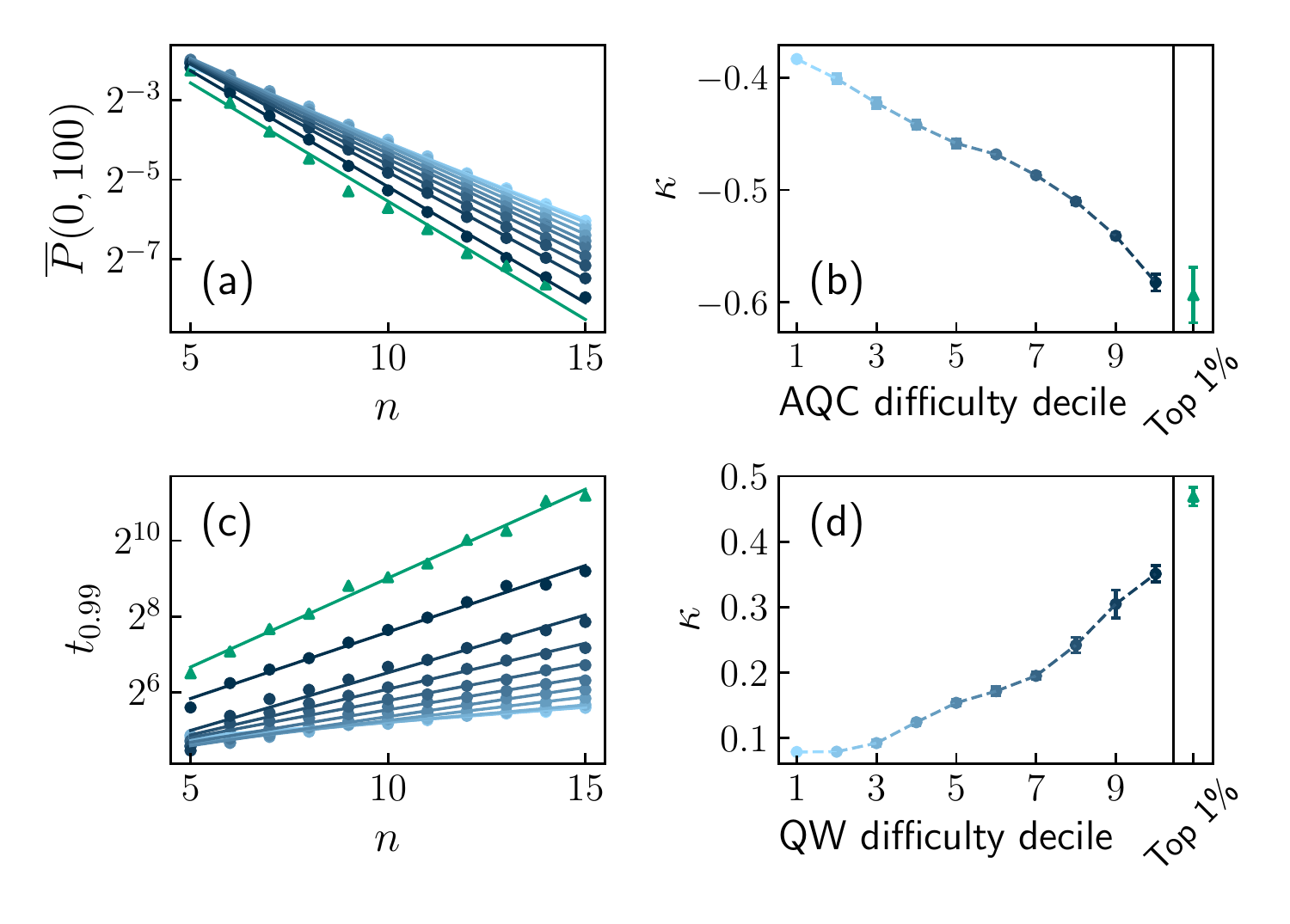}
    \caption{
    (a) The median value of the QW success probability $\overline{P}(0, 100)$ of each of the AQC difficulty deciles (blue circles) and the most difficult 1\% of instances for AQC (green triangles) plotted against the number of variables $n$ on log-linear axes, with a linear fit (solid line) for each category. Darker shades of blue represent more difficult deciles. (b) Plot of the values of the scaling exponents $\kappa$ that have been inferred from the gradients of the linear fits, with error bars indicating standard errors. (c) and (d) Similar plots for the median values of the AQC duration $t_{0.99}$ for instances organised by QW difficulty. Note that since there were more than 100 instances that we could not calculate $t_{0.99}$ for at $n=15$, the plot in (a) does not include a point at $n=15$ for the median of $\overline{P}(0, 100)$ for the most difficult 1\% of instances for AQC.
    }
    \label{fig:qw and aqc deciles cross comparison}
\end{figure}

To make a cross-comparison between QW and AQC, we examine the QW difficulty of instances when grouped by AQC difficulty, and vice versa. In Fig.~\ref{fig:qw and aqc deciles cross comparison}(a), we plot the median values of the QW success probability $\overline{P}(0, 100)$ for the instances in each AQC difficulty decile against $n$ on log-linear axes. The median values of $\overline{P}(0, 100)$ for the most difficult 1\% of instances for AQC are also plotted. A straight line is fit to the points for each of these groups of instances, and the corresponding scaling exponents are plotted in Fig.~\ref{fig:qw and aqc deciles cross comparison}(b). The fact that more difficult AQC deciles tend to correspond to smaller median values of $\overline{P}(0, 100)$ implies that AQC difficulty is a good indicator of QW difficulty. Note that this is not the case at the lowest values of $n$, but the correlation becomes more clear as $n$ is increased. This agrees with the correlations we found in Fig.~\ref{fig:m2saqcqwhexbin}. The implied scaling exponents show that this increase in QW difficulty with AQC difficulty becomes more prominent across all of the deciles as $n$ is increased. In particular, the correlation between QW difficulty and AQC difficulty seems to remain strong at the ``difficult tail'' of the distribution. However, there isn't a large jump in the scaling for the most difficult 1\% of instances for AQC on QW, which shows that the extremely difficult instances for AQC aren't as extremely difficult for QW. This indicates a significant advantage of the portfolio approach for solving the instances that AQC finds extremely difficult.

We make a similar comparison in Figs.~\ref{fig:qw and aqc deciles cross comparison}(c) and~\ref{fig:qw and aqc deciles cross comparison}(d), where we plot the median AQC durations $t_{0.99}$, and the corresponding scaling exponents, for the instances in QW difficulty deciles and the 99th percentile instances for QW difficulty. In accordance with the previous results, these plots indicate that QW difficulty is a good indicator of AQC difficulty, though this is again more clearly the case for larger $n$. The trend of worsening scaling exponent with increasing difficulty decile in both sets of comparisons indicates that the QW-AQC difficulty correlation is likely to continue to larger $n$, even for the tail of difficult instances. Therefore, a simple portfolio-based strategy is unlikely to produce a drastic speedup, though there is still room for some speedup, especially for the instances that are extremely difficult for AQC. Assuming that performing runs of both QW and AQC for each instance incurs a cost of roughly a factor of two to the total run time, even a small scaling advantage obtained from this approach would make up for this cost at large problem sizes.

\subsection{\label{sec:quantum/classical_difficulty}Quantum/classical difficulty comparison}

To quantify the classical difficulty of MAX 2-SAT instances, we have measured the number of problem calls $N_\mathrm{calls}$ made by the classical algorithm MixBandB when solving each MAX 2-SAT instance. While there exist many classical algorithms that perform much better than MixBandB, they typically employ heuristic methods to gain a speed advantage, which make a significant difference at small problem sizes. This is undesirable for our analysis because the quantum algorithms we are comparing them to do not use such heuristics.
MixBandB is based on MIXSAT, which is a powerful MAX SAT solver, but it does not incorporate the heuristic optimizations that MIXSAT uses.
In this sense, MixBandB is a good classical comparison to QW and AQC. However, MixBandB is an exact solver, meaning that it always returns an optimal solution at the end of a run, whereas QW and AQC cannot give such guarantees. Since the purpose of this work is not to benchmark the scalings of these algorithms but to compare the underlying mechanisms they use to solve problems, this difference is not important for our analysis.

\begin{figure}
\centering
    \includegraphics[width=\columnwidth]{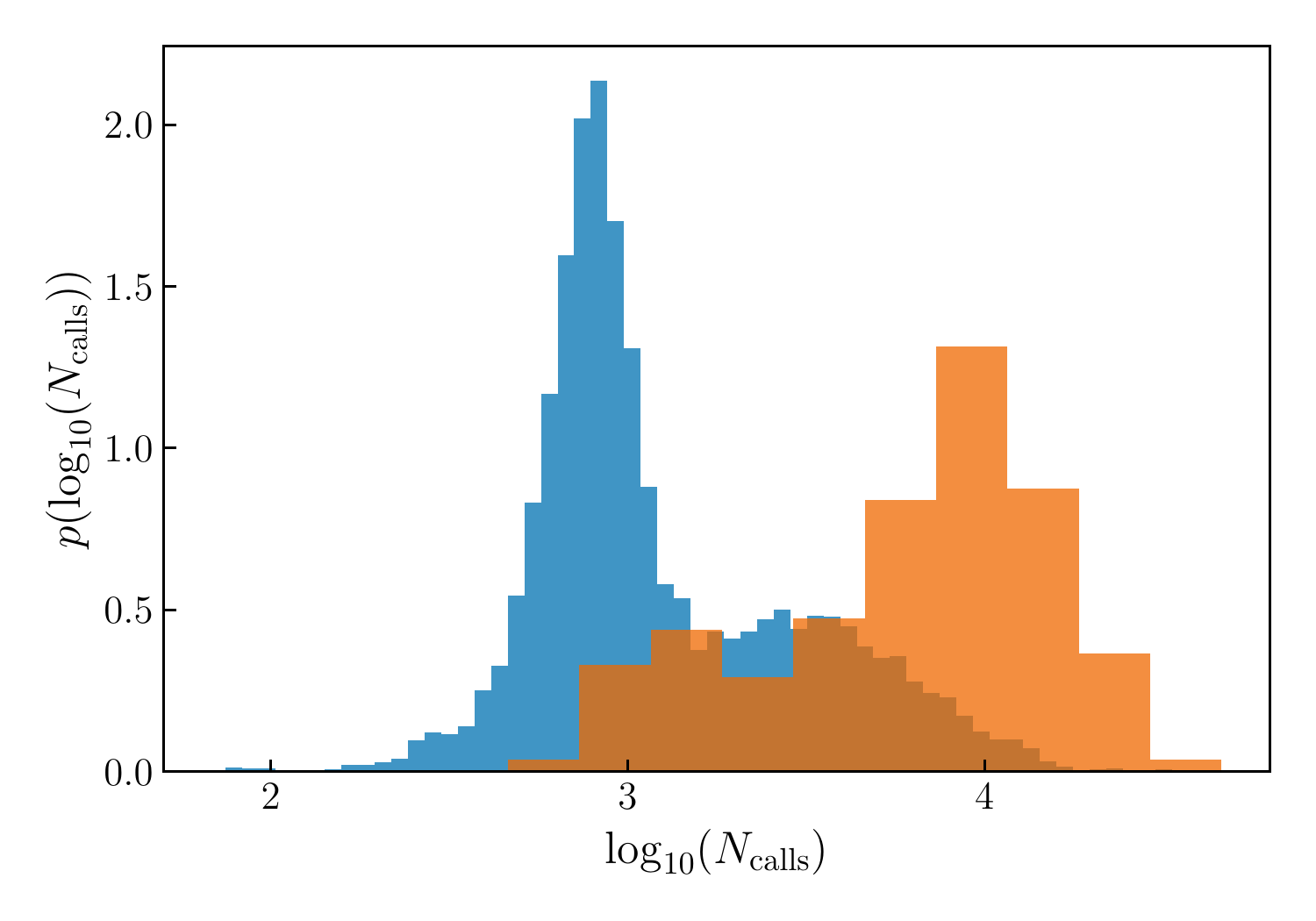}
    \caption{
    Histogram showing the approximate probability density $p(\log_{10}(N_\mathrm{calls}))$ of the logarithm of the number of problem calls $N_\mathrm{calls}$ made by the MixBandB algorithm when solving the typical $n=20$ instances (blue) and the instances from~\cite{Crosson2014} (orange), which are difficult for QA.}
    \label{fig:m2s_mixbnb20hist}
\end{figure}

Fig.~\ref{fig:m2s_mixbnb20hist} shows a histogram of the approximate probability density of the logarithm of the number of calls $N_\mathrm{calls}$ made by MixBandB for typical instances with $n=20$ variables and the instances that are difficult for QA. It can be seen that the typical instances form a bimodal distribution, where there is a peak of instances that required relatively few calls and a long tail of more difficult instances that form another peak at a higher number of calls. This suggests that the typical instances can be roughly divided into two sets for their difficulty classically, with the majority of instances being in the less difficult set. The instances that are difficult for QA also form a bimodal distribution, but with a larger peak of difficult instances than less difficult instances. The centre of this distribution is located on the ``difficult tail'' of typical instances, suggesting that difficult instances for QA tend to be difficult for MixBandB too.

\begin{figure}
\centering
    \includegraphics[width=\columnwidth]{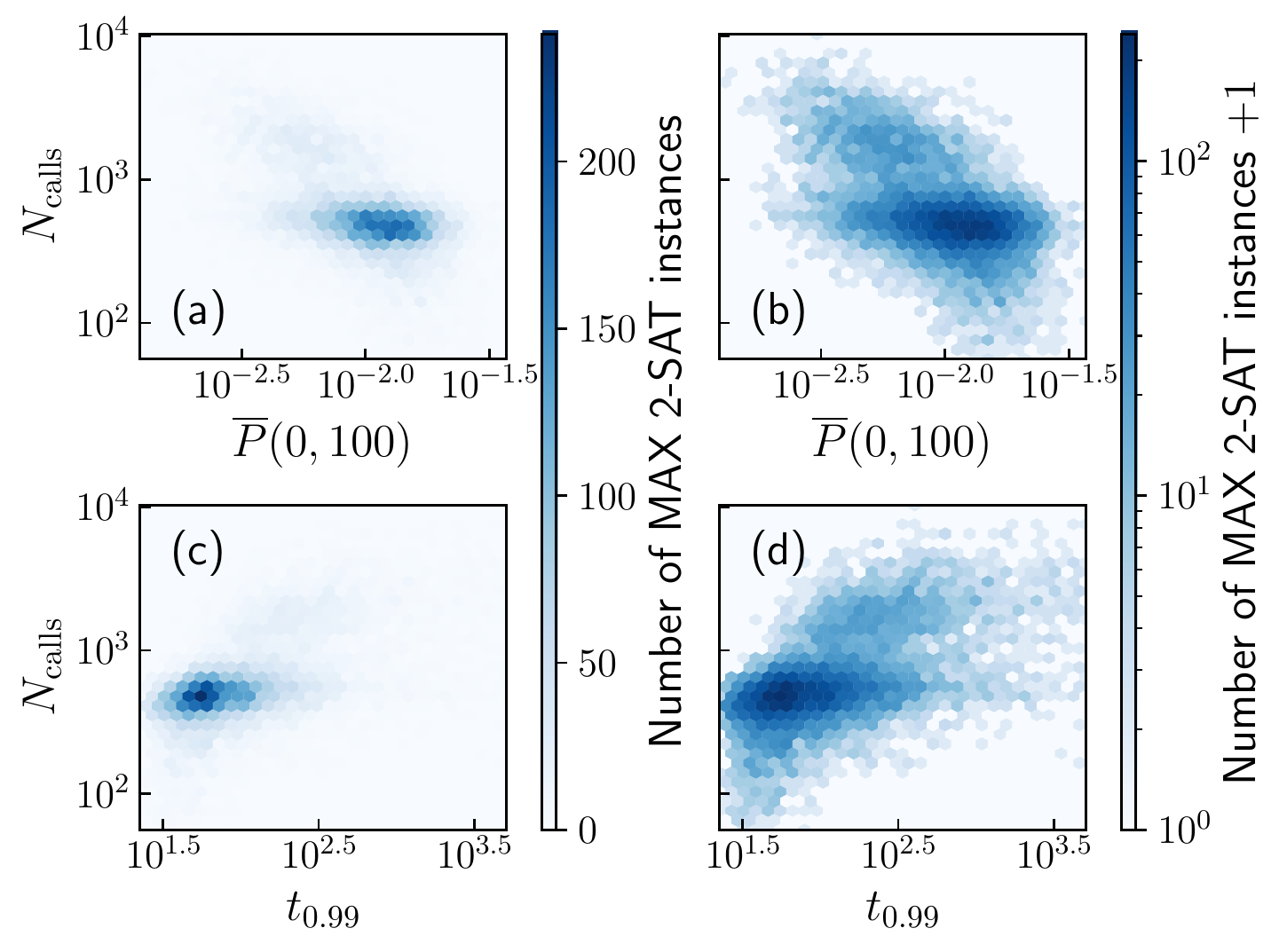}
    \caption{Heatmaps of the number of problem calls $N_\mathrm{calls}$ against (a) and (b) the QW success probability $\overline{P}(0,100)$ and (c) and (d) the AQC duration $t_{0.99}$ on log-log axes for typical instances with $n=15$ variables. 138 instances for which $t_{0.99}$ was not successfully calculated are excluded in (c) and (d). (a) and (c) show the heatmaps with a linear colour scale, and (b) and (d) show the same distributions with a logarithmic colour scale, where the value for each cell has been increased by one to remove the zeros. Visually, some correlation can be seen in both distributions. The Spearman's rank correlation coefficients are $\approx -0.47$ for the distribution in (a) and $\approx 0.52$ for the distribution in (c).
    \label{fig:m2saqcqwmixbnbhexbins} 
    }
\end{figure}

In Fig.~\ref{fig:m2saqcqwmixbnbhexbins}, we plot the joint distribution of $N_\mathrm{calls}$ and $\overline{P}(0,100)$ for the typical instances with $n=15$ variables, and the joint distribution of $N_\mathrm{calls}$ and $t_{0.99}$ for the 9862 typical instances that we successfully calculated $t_{0.99}$ for. The former distribution has a Spearman's rank correlation coefficient of $\approx -0.47$ and the latter has a Spearman's rank correlation coefficient of $\approx 0.52$. These imply that correlations exist between difficulty for MixBandB and difficulty for both QW and AQC, although the correlations are not as strong as what we previously observed between QW difficulty and AQC difficulty at the same problem size. Therefore, it seems that a portfolio-based strategy would be more effective when applied to a quantum and classical algorithm, as opposed to QW and AQC. This does not rule out the possibility of attaining a better performance from other forms of (hybrid) quantum algorithms.
Examples of other techniques in continuous-time quantum computing that can be incorporated in a hybrid approach include: pre-annealing~\cite{callison2021energetic}; local quantum searches, which can be performed with the addition of a biased Hamiltonian in coherent annealing~\cite{Perdomo-Ortiz2011, Duan2013, ohkuwa2018reverse, Grass2019} or with just the driver and problem Hamiltonians in dissipative reverse annealing~\cite{chancellor2017modernizing}; annealing schedules that interpolate between QW and AQC~\cite{Morley2017}; and a variety of approaches that fall under the umbrella of diabatic quantum computing, which are reviewed in \cite{crosson2021prospects}. For a detailed review of hybrid approaches involving quantum and classical algorithms, see~\cite{Callison2022}.

We note that since different classical algorithms will have varying levels of correlation with each other, our measure of ``classical difficulty'' cannot be extended to represent difficulty for classical algorithms as a whole, since such a thing does not exist. However, the fact that some correlation exists between difficulty for the quantum algorithms and difficulty for MixBandB is still interesting, as it indicates that there are some characteristics of instances that make them typically more difficult for both the quantum algorithms and some ``good'' classical algorithms. Further work is required to find out if there are other good classical algorithms with lower levels of correlation with the quantum algorithms.

\subsection{\label{sec:satisfiable_instances}Satisfiable instances}

Satisfiable instances of MAX 2-SAT are easy to solve classically because the optimal solution can be found with a 2-SAT solver, and 2-SAT is known to be in P; a linear time algorithm for 2-SAT was found in~\cite{Aspvall1979}, which is based on finding the strongly connected components of the problem's implication graph. Some of the typical instances that we are analysing are satisfiable, the proportion of which decreases with $n$. A good classical MAX SAT solver will take advantage of this to solve satisfiable instances efficiently, for example by calling a 2-SAT solver at the start of the algorithm. However, since the quantum algorithms we are studying don't explicitly check for the satisfiability of formulae, it is unclear whether they will solve satisfiable instances efficiently. In this subsection, we analyse the difficulty of satisfiable instances for QW and AQC and discuss the implications of this for a portfolio approach.

\begin{figure}
    \includegraphics[width=\columnwidth]{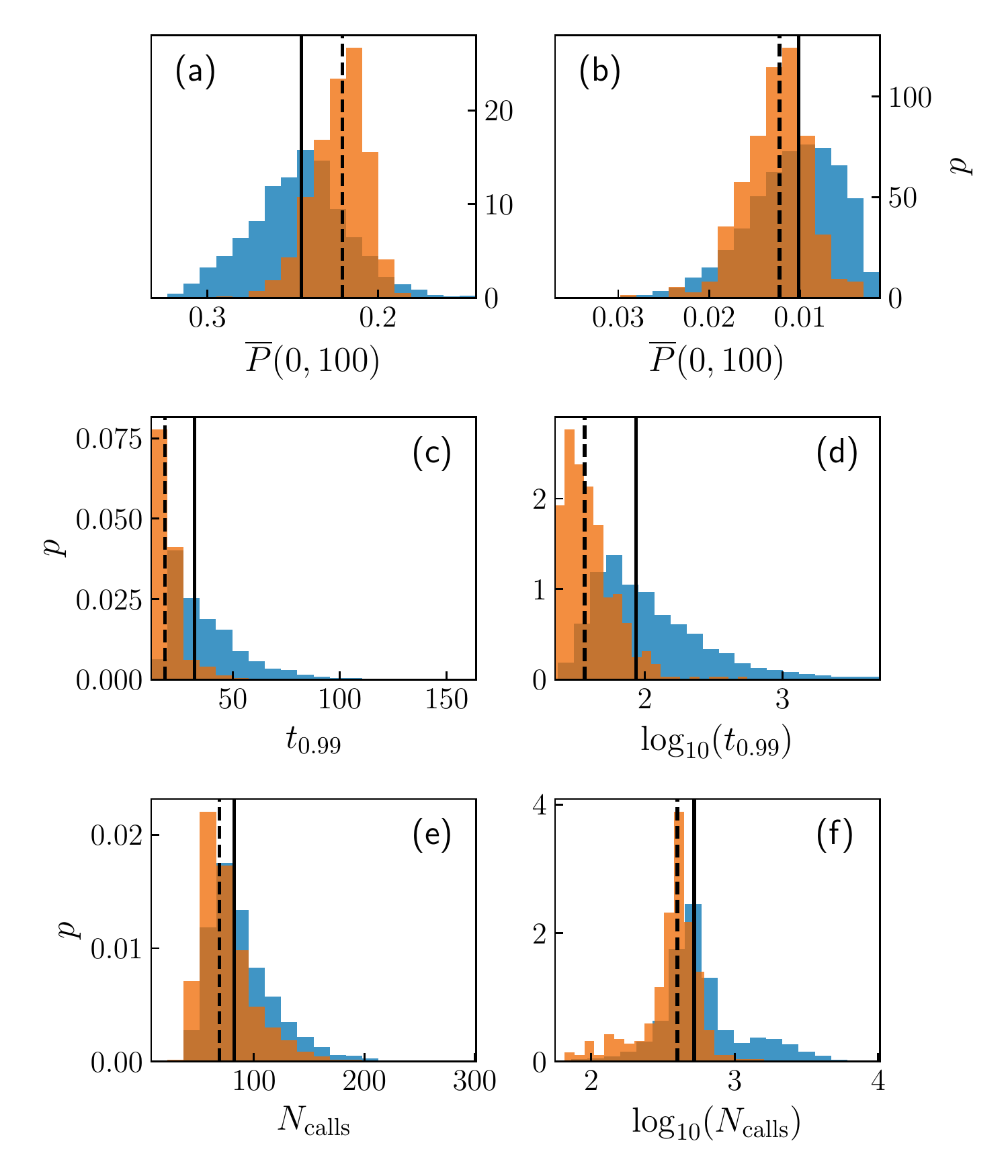}
    \caption{Histograms of the approximate probability density $p$ of (a) and (b) the QW success probability $\overline{P}(0, 100)$, (c) AQC duration $t_{0.99}$, (d) logarithm of $t_{0.99}$, (e) MixBandB calls $N_{\mathrm{calls}}$, and (f) logarithm of $N_{\mathrm{calls}}$ for typical satisfiable (orange) and unsatisfiable (blue) MAX 2-SAT instances with (a), (c), and (e) $n=5$ variables and (b), (d), and (f) $n=15$ variables. The x-axes have been reversed in (a) and (b) so that instance difficulty increases from left to right for all plots. Median values of the distributions are indicated by dashed lines for satisfiable instances and solid lines for unsatisfiable instances. Note that the distributions in (d) do not include 138 instances for which $t_{0.99}$ was not successfully calculated.
    }
    \label{histograms satisfiable vs unsatisfiable}
\end{figure}

The plots in Figs.~\ref{histograms satisfiable vs unsatisfiable}(a) and~\ref{histograms satisfiable vs unsatisfiable}(b) show the approximate probability density of the QW success probability $\overline{P}(0, 100)$ for satisfiable and unsatisfiable instances with $n=5$ and $n=15$ variables. The median value of each distribution is indicated by a vertical line. We find that the median $\overline{P}(0, 100)$ for satisfiable instances is larger than for unsatisfiable instances at $n=15$, but at $n=5$ the satisfiable instances have a smaller median $\overline{P}(0, 100)$ than the unsatisfiable instances. This indicates that QW finds satisfiable instances less difficult on average except in the case of very small $n$, although even at $n=5$ the most difficult instances still tend to be unsatisfiable. We note that the long tail of difficult instances cannot be as easily identified in Fig.~\ref{histograms satisfiable vs unsatisfiable}(b) as it can in Fig.~\ref{fig:m2saqcqwhexbin}, which is because we are using a linear x-axis in Fig.~\ref{histograms satisfiable vs unsatisfiable} and QW difficulty is inversely proportional to $\overline{P}(0, 100)$. The tail of difficult instances can be clearly seen in both of the $n=15$ distributions when plotting $1/\overline{P}(0, 100)$ or using a logarithmic x-axis.

In practice, the satisfiable instances are less difficult for QW than what can be inferred from the plots in Fig.~\ref{histograms satisfiable vs unsatisfiable}. This is due to the fact that the QW success probability is generally much less than 1, so repeat runs are needed to increase the probability of finding the optimal solution. From Eq.~\eqref{eq:clause_hamiltonian}, we can see that each unsatisfied clause adds an energy contribution of 1. A simple improvement to a QW strategy would be to measure the energy of the final state or classically evaluate the formula with the solution given by QW to efficiently determine the number of clauses that are left unsatisfied by the corresponding solution. If the optimal solution of a satisfiable instance is found, it would become immediately obvious that there are no more clauses that can be satisfied, so repeat runs would no longer be necessary. For unsatisfiable instances, we cannot be certain that we have found an optimal solution based on the information gained from QW alone, so there will typically be ``wasted'' runs conducted after finding the optimal solution. The number of extra runs depends on the specific strategy used to determine when to stop doing repeats. There has been previous work on applying sophisticated methods of determining the stopping point to quantum annealing \cite{Vinci2016a}.

Figs.~\ref{histograms satisfiable vs unsatisfiable}(c) and~\ref{histograms satisfiable vs unsatisfiable}(d) show the approximate probability density of the AQC duration $t_{0.99}$ for typical satisfiable and unsatisfiable instances with $n=5$ variables, and similarly for the approximate probability density of the logarithm of $t_{0.99}$ for $n=15$. The satisfiable instances have shorter median durations than the unsatisfiable instances for both $n=5$ and $n=15$, indicating that AQC finds these problems less difficult on average. Since AQC achieves a high success probability in a single run, we do not need to do repeat runs as in the case for QW, so these plots are a good representation of the difference in AQC difficulty between satisfiable and unsatisfiable instances. Figs.~\ref{histograms satisfiable vs unsatisfiable}(e) and~\ref{histograms satisfiable vs unsatisfiable}(f) show similar results for the MixBandB algorithm, which like the quantum algorithms does not check for satisfiability. Satisfiable instances are found less difficult on average, but the distributions for satisfiable and unsatisfiable instances overlap.

The significant overlap between the distributions of satisfiable and unsatisfiable instances in Fig.~\ref{histograms satisfiable vs unsatisfiable} either imply that QW and AQC do not find the difficulty of satisfiable instances to be as low as their difficulty for the best classical algorithms, or that a large fraction of the unsatisfiable instances at these problem sizes have just as low difficulties as the satisfiable instances, which can be solved efficiently. To determine which of these two cases is true, we can check whether the difficulty of satisfiable instances scales exponentially for the quantum algorithms, which would support the former as it would indicate that QW and AQC cannot solve them efficiently.

\begin{figure}
    \includegraphics[width=\columnwidth]{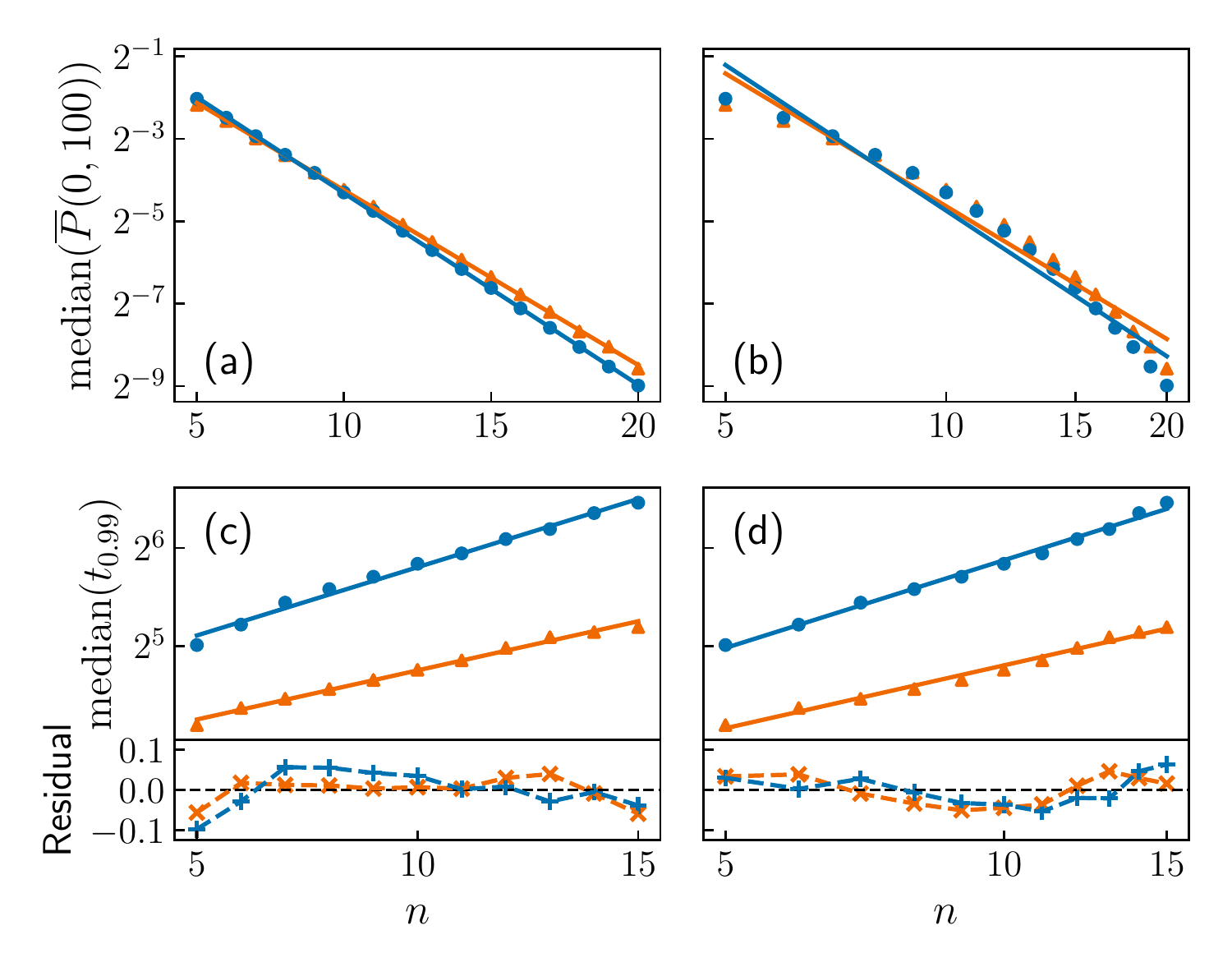}
    \caption{Median QW success probability $\overline{P}(0, 100)$ for all satisfiable (orange triangles) and unsatisfiable (blue circles) typical instances against the number of variables $n$ on (a) log-linear and (b) log-log axes. Linear fits are shown for both the satisfiable and unsatisfiable instances. (c) and (d) The same as above, but for the AQC duration $t_{0.99}$. Residuals, which are calculated logarithmically as $\log_2(\mathrm{median}(t_{0.99})) - \log_2(y(n))$, where $y(n)$ is the corresponding fit, are also shown in (c) and (d) for both the satisfiable (orange crosses) and unsatisfiable (blue pluses) instances.
    }
    \label{scalings satisfiable vs unsatisfiable}
\end{figure}

We plot the scaling of the median QW success probability and AQC duration with $n$ for satisfiable and unsatisfiable instances on log-linear and log-log axes in Fig.~\ref{scalings satisfiable vs unsatisfiable}. An exponential scaling would fit better to a straight line on log-linear axes, whereas a polynomial scaling would have a better linear fit on log-log axes. For QW, we can see that the log-linear axes produce better fits for both sets of instances, meaning that the median of $\overline{P}(0, 100)$ appears to scale exponentially with $n$ for both satisfiable and unsatisfiable instances. For AQC, it is unclear which fits are better. We note that these results cannot be used to make statements about the form of the scalings with certainty, as the problem sizes are very small compared to practically relevant instances and the scalings may change at larger sizes.

Given the linear worst-case scaling of good classical 2-SAT solvers and the apparent exponential scaling of QW (and potentially AQC) on satisfiable instances, we can conclude that a classical algorithm should most likely be used to efficiently check the satisfiability of instances at the start of any portfolio-based strategy involving QW and/or AQC, as these instances cannot be solved efficiently by the quantum algorithms. This will speed up the time to solution for satisfiable instances while only contributing a linear overhead to the run time for unsatisfiable instances. An added benefit of this approach is that instances with exactly one clause left unsatisfied by an optimal assignment can be solved faster by QW. This is because by running a classical 2-SAT solver in advance, it would become known when a formula is unsatisfiable and that any assignment satisfying all but one clause is optimal. Therefore, there would be no need for repeat runs of QW after an optimal solution is found, for the same reasons as we mentioned for satisfiable instances. This does not apply to AQC as it does not require repeat runs.

\section{\label{sec:conclusions}Conclusions}

We have examined both the relative difficulty of different instances and the correlation in difficulty for a selection of both classical and quantum algorithms. These include algorithms that behave in conceptually different ways, for example quantum walk, which relies on many repeats with a relatively low success probability, versus adiabatic quantum computing, which succeeds with a high probability after a single run. Our work shows that it is important to include a thorough characterisation of the problem instances used for numerical studies of the performance of quantum algorithms. We have found that while there is some correlation in MAX 2-SAT instance difficulty between methods, the correlation seems weak enough that a strategy of attempting a portfolio of algorithms in parallel is viable and likely to be desirable in real computation. We also note unique features of specific strategies. For example, the performance of the most difficult instances for AQC is drastically worse than for more ``typical'' problems, much more so than for quantum walk. This can be attributed to the presence of instances with extremely small spectral gaps, which limit the performance of AQC~\cite{Knysh2016, Albash2018}. Extremely small gaps have been observed in other contexts~\cite{Chancellor2021}. This catastrophic failure of AQC suggests that a ``stand-alone'' adiabatic strategy without attempting others in parallel is likely to be particularly undesirable. We further find that while the performance of quantum algorithms is generally better for satisfiable problems (which can be solved efficiently classically), these problems are still not solved efficiently by either of the quantum algorithms. This strongly suggests that first performing a classical check for satisfiability is useful. In a sense, attempting different algorithms in parallel can be seen as the most trivial case of a hybrid algorithm. If the algorithms are classical and quantum then it is a hybrid quantum-classical algorithm, but there can also be hybrids between two quantum algorithms, such as AQC and QW. While not the topic of this paper, the fact that even such simple hybrid methods are desirable bodes well for more complicated methods of combining algorithms that are likely to lead to further gains, for example pre-annealing in \cite{callison2021energetic} as a quantum-quantum algorithm, or various biasing and reverse annealing techniques \cite{chancellor2017modernizing} as examples of quantum-classical hybrids.

The correlations do suggest that while attempting multiple different algorithms in parallel is likely to be fruitful, there is also likely a more fundamental sense of difficulty in terms of being resistant to being efficiently solved by any algorithm, quantum or classical. Furthermore, the double peaked nature of many of the distributions of effort required for problems suggests that the transition toward being predominantly difficult as problems scale toward the large size limit is not simple, at least not in the case of MAX 2-SAT. This behaviour is partially explained by the difference between satisfiable and unsatisfiable instances, but this appears not to be the whole story because there is significant overlap between the two in terms of difficulty. While we have made significant steps in understanding relative problem difficulty over different algorithms at sizes relevant for exhaustively simulated quantum computing, there is still much work to be done to fully understand this important topic which underlies many numerical studies.

\vspace{5mm}

The data for all MAX 2-SAT problem instances used in this
research are openly available at~\cite{Callison2023data}.

\begin{acknowledgments}
We thank Elizabeth Crosson for providing data for difficult MAX 2-SAT instances. We also thank the members of Algorithms and Complexity in Durham for helpful discussions.
NC and VK were supported by EP/L022303/1 and impact acceleration funding associated with this grant.
NC was supported by EPSRC fellowship EP/S00114X/1.
PM was supported by EPSRC Doctoral Training Funds awarded to Durham University (EP/T518001/1) in partnership with dunnhumby.
AC was funded by UKRI EPSRC Grant No. EP/L016524/1 via
the Imperial College London Centre for Doctoral Training in Controlled Quantum Dynamics, and the EPSRC UK Quantum Technology Hub in Computing and Simulation (EP/T001062/1).
\end{acknowledgments}

\appendix
\setcounter{secnumdepth}{0}
\section{Appendix}

\begin{figure}
    \centering
    \includegraphics[width=\columnwidth]{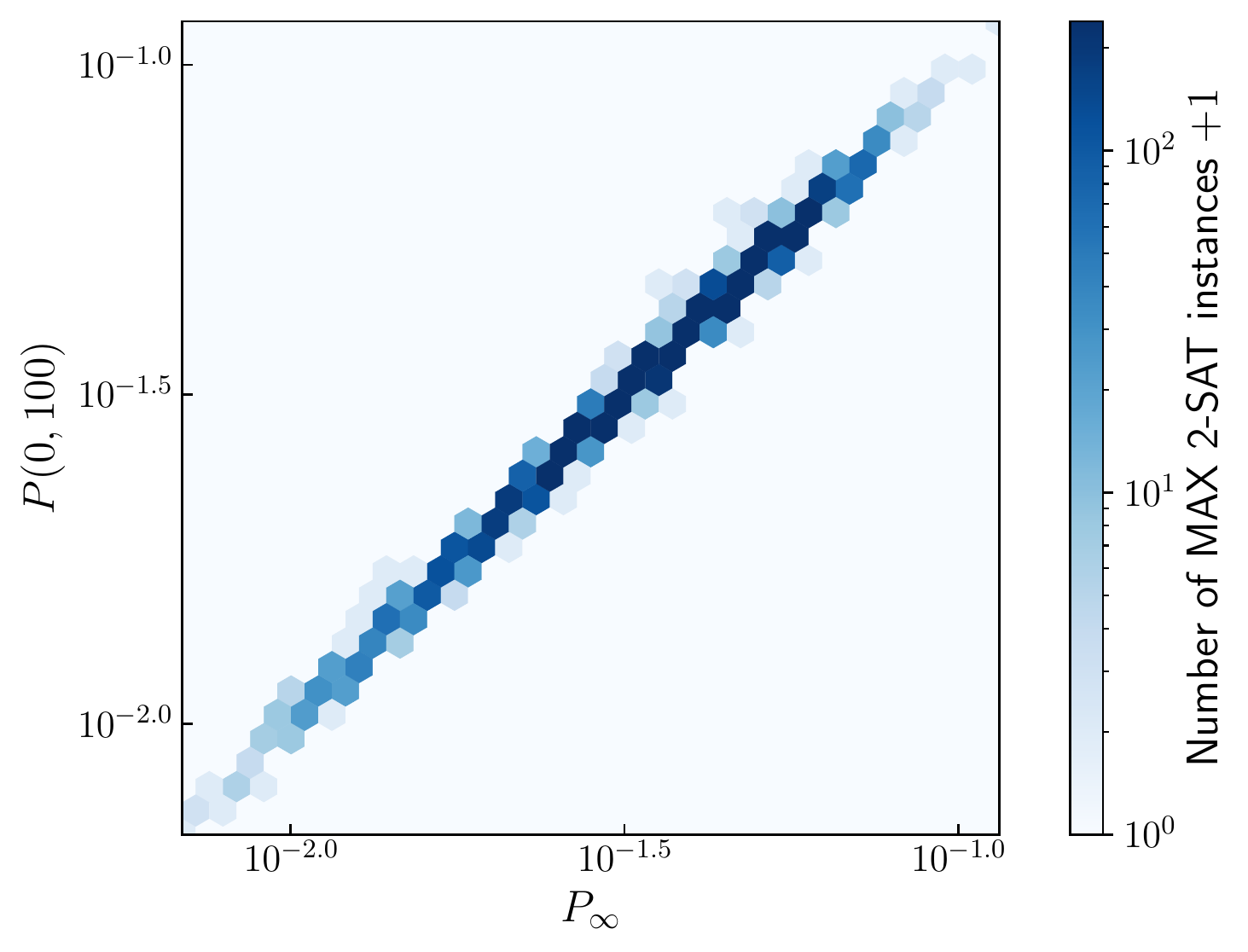}
    \caption{
    Heatmap of the QW success probability $\overline{P}(0, 100)$ averaged over a time interval $\Delta t_I = 100$ against the same average probability in the limit of infinite time interval, $P_\infty$, for the typical MAX 2-SAT instances with $n=11$ variables. A logarithmic colour scale is used for better visibility, and the zeroes have been removed by adding 1 to the number of instances for each cell. The Spearman's rank correlation coefficient between the two quantities is $\approx 1.00$, indicating an almost perfect correlation.
    }
    \label{fig:m2sfiniteinfinitetimeprobs}
\end{figure}

Our numerical analysis of QW is based on the average success probability $\overline{P}(0, 100)$, which is taken over an arbitrary time interval $\Delta t_I = 100$. Since this time interval is longer than the timescale of the QW dynamics (see Fig.~\ref{fig:m2sinstsuccessprobs}), we expect that the success probability would not be significantly affected by a different choice of $\Delta t_I$ that is also sufficiently large. To confirm this, we consider the infinite time interval limit of the average success probability, $P_\infty \equiv \lim_{\Delta t_I \to \infty} \overline{P}(0, \Delta t_I)$. We followed the procedure outlined in~\cite{Callison2019} to calculate $P_\infty$ by numerically diagonalising the QW Hamiltonian. Fig.~\ref{fig:m2sfiniteinfinitetimeprobs} shows that for the typical instances with $n=11$ variables, $P_\infty$ and $\overline{P}(0, 100)$ are in very good agreement. Therefore, we can assume that the results of this paper are not dependent on our specific choice of $\Delta t_I$.

\end{document}